\documentclass[a4paper]{article}

\usepackage{icrc2013}

\title{Towards a full Atmospheric Calibration system for the Cherenkov
  Telescope Array}

\shorttitle{Atmospheric Calibration for CTA}

\authors{
M.~Doro$^{1,2,3}$,
M.~Gaug$^{2,3}$,
O.~Blanch$^{4}$,
Ll. Font$^{2,3}$,
D. Garrido$^{2,3}$, 
A. L\'opez-Oramas$^{4}$, 
M. Mart\'inez$^{4}$ for the CTA consortium 
}

\afiliations{
$^1$ University and INFN Padova, via Marzolo 8, 35131 Padova (Italy) \\
$^2$ F{\'i}sica de les Radiacions, Departament de F{\'i}sica,
  Universitat Aut{\`o}noma de Barcelona, 08193 Bellaterra, Spain \\ 
$^3$  CERES, Universitat Aut{\`o}noma de Barcelona-IEEC, 08193 Bellaterra,
  Spain \\ 
$^4$ Institut de Fis{\`i}ca d'Altes Energies, 08193 Bellaterra, Spain 
}

\email{michele.doro@pd.infn.it}

\abstract{The current generation of Cherenkov telescopes is mainly
  limited in their gamma-ray energy and flux reconstruction by
  uncertainties in the determination of atmospheric parameters. The
  Cherenkov Telescope Array (CTA) aims to provide high-precision data
  extending the duty cycle as much as possible. To reach this
  goal, it is necessary to continuously and precisely monitor the
  atmosphere by means of remote-sensing devices, which are able to
  provide altitude-resolved and wavelength-dependent extinction
  factors, sensitive up to the tropopause and higher. Raman LIDARs are
  currently the best suited technology to achieve this goal with one
  single instrument. However, the synergy with other instruments
  like radiometers, solar and stellar photometers, all-sky cameras,
  and possibly radio-sondes is desirable in order to provide more
  precise and accurate results, and allows for weather forecasts and
  now-casts. In this contribution, we will discuss the need and
  features of such multifaceted atmospheric calibration systems.} 


\begin{document}
\maketitle

\section{Introduction}
Currently in its design stage, the Cherenkov Telescope Array (CTA) is
an advanced facility for ground-based very-high-energy  gamma-ray
astronomy~\cite{bib:actis}. It is an international initiative to
build the next-generation Cherenkov telescope array covering the
energy range from a few tens of GeV to a few hundreds of TeV with an
unprecedented sensitivity. The design of CTA is based on currently
available technologies and builds upon the success of the present
generation of ground-based Cherenkov telescope arrays (H.E.S.S., MAGIC
and VERITAS\,\footnote{\url{www.mpi-hd.mpg.de/hfm/HESS/},\\
  \url{wwwmagic. mppmu.mpg.de}, \\\url{veritas.sao.arizona.edu}}).  

Nowadays, the main contribution to the systematic uncertainties  
of imaging Cherenkov telescopes stems from the uncertainty in the
height- and wavelength-dependent atmospheric transmission for a
given run of data. MAGIC cites a contribution of 10\% to the
uncertainty of their energy scale~\cite{bib:aleksic} and 12\%
additional uncertainty on the flux due to run-by-run variations, while
H.E.S.S. retrieves 10\% for the atmospheric profile, and 15\% from
run-by-run atmospheric variations~\cite{bib:aharonian}. Both
estimates are based upon data recorded during clean atmospheric
conditions and have to be considered lower-limits for the general case
of data taken under moderately acceptable atmospheric
conditions. Atmospheric quality affects the measured Cherenkov yield
in several ways: The air-shower development itself, the loss of
photons due to scattering and absorption of Cherenkov light out of
the camera field-of-view, resulting in dimmer images and the
scattering of photons into the camera, resulting in blurred images. 
Despite the fact that several supplementary instruments are currently
used to measure the atmospheric transparency, their data are only used
to retain good-quality observation time slots, and only a minor effort
has been made to routinely correct data with atmospheric
information~\cite{bib:nolan1,bib:dorner,bib:reyes}.  

It is envisaged that the world-wide community of scientists using CTA
data will be serviced with high-level data. It is moreover foreseeable
that CTA will observe many more spectral features than the current
generation of Imaging Atmospheric Cherenkov Telescopes (IACTs), probably also resolving finer structures. To 
achieve this goal, the atmosphere must be monitored continuously and
precisely such that observatory data can be corrected before
dissemination. This requires the extensive use of remote-sensing
instrumentation such as LIDARs, possibly complemented by additional
atmospheric monitoring devices to complement the LIDAR information. 

\vspace{-5mm}
\section{Effects of atmospheres on data reconstruction}
Although IACTs are normally placed at astronomical sites,
characterized by extremely good atmospheric conditions, the local
atmosphere is potentially influenced by phenomena occurring at tens to
thousands of kilometers far, and thus should be continuously
monitored. Of the various atmospheric layers, only the
troposphere (reaching up to $\sim$15~km) and sometimes parts of the
tropopause and, in the case of stratovolcanic eruptions, the lower
stratosphere (15--20~km)  are affected by variations of their chemical
(and thus optical) properties. Air molecules can travel to the top of the troposphere
(from 7 to 20~km depending on the latitude) and back down in a few
days, hence this mixing makes the characteristics of the layer
changing fast.  
%
%
While the molecular content of the atmosphere varies very slowly at a
given location during the year, and slowly from place to place,
aerosol concentrations can vary on time-scales of minutes and travel
large, inter-continental, distances. Most of them are concentrated
within the first 3~km of the troposphere, with the
free troposphere above being orders of magnitude cleaner. 
%
Aerosol sizes reach from molecular dimensions to millimeters, and the particles
remain in the troposphere from 10 days to 3 weeks. The sizes are
strongly dependent on relative humidity. 
Different types of aerosol show characteristic size distributions, and
an astronomical site 
will always show a mixture of types, with one possibly dominant type at a
given time and/or altitude.
%
%
Light scattering and absorption by aerosols needs to be described by
Mie theory or further developments of it, including  non-sphericity of
the scatterer. Aerosols use to have larger refraction indices than the
one of water,  and typically show also a small imaginary
part. Contrary to the typical $\lambda^{-4}$ wavelength dependency of
Rayleigh-scattering molecules, aerosols show power-law indices (the
so-called \textit{\AA ngstr\"om} coefficients) from  0 to 1.5, i.e. a
much weaker dependency on wavelength. 

In order to estimate the effect of different atmospheric conditions on
the image analysis of IACTs, we have simulated different molecular and
aerosol profiles for the MAGIC system, consisting of two
telescopes. The results are presented elsewhere in this
conference~\cite{bib:garrido}. Several aerosol scenarios have then
been simulated: Enhancements of the ground layer from a quasi
aerosol-free case up to  a thick layer which reduces optical
transmission by 70\%, a cloud layer at the altitudes of 6~km,
10~km (cirrus)  and 14~km~(volcano debris)~a.s.l.  and a 6~km
cloud layer with varying aerosol densities. We found --- 
as expected --- that the aerosol and clouds layer height,
besides the density and type, affect the data differently, and that
the position of this overdensity should be known precisely. In
other words, the \emph{total} extinction (or the Aerosol Optical
Depth) is not a good parameter for all cases, and using only integral
extinction often may lead to large systematic errors. For this reason,
height-resolving instruments are required. We believe that the main findings of this study should also
be valid for CTA, at least in the energy range from 50 GeV to 50 TeV,
albeit efforts have started to repeat the same simulations for
CTA. Previous studies have been made
\cite{bib:bernloher,bib:nolan1,bib:dorner} for H.E.S.S. and
for the MAGIC mono system, however only for an increase of low-altitude
aerosol densities, and in \cite{bib:nolan2} for a reference
configuration of CTA, claiming a change in the spectral power-law
index of  gamma-ray fluxes, when atmospheric aerosol layers are
present. In our work, we found that different atmospheres affect the
energy threshold, the energy resolution and the 
energy bias, that propagate into the computation of a target flux and
spectral reconstruction. See \cite{bib:garrido} for further details. 
\vspace{-5mm}
\section{Raman LIDARs for CTA}
Atmospheric properties can be derived, to a certain extent, directly
from IACT data.  Several studies have been made by the H.E.S.S. and
MAGIC collaborations to estimate the integral atmospheric
transmission, using trigger rates, muon rates, combinations of both~\cite{bib:reyes},
or the anode currents of the photomultipliers and/or pedestal RMS. Up
to now, these parameters have been used only to discard data taken
under non-optimal conditions, but work is ongoing to use this
information to correct data themselves. However, as stated above, the
use of integral transmission parameters is only valid in some of the
possible atmospheric scenarios, namely those where the aerosol
enhancement is found at the ground layer, or where clouds are low (below few
km a.g.l), since the integral transmission parameters lack information
about the layer height. For layers at higher altitudes, trigger rates
with different cuts in image size and the stereo shower parameters
themselves  could eventually be used, however studies on these
possibilities are not yet conclusive. For this reason, 
we have investigated the possibilities of using remote sensing devices
such as the LIDAR \cite{bib:lopez}.  

\begin{figure}[h!t]
\centering
\includegraphics[width=0.8\linewidth]{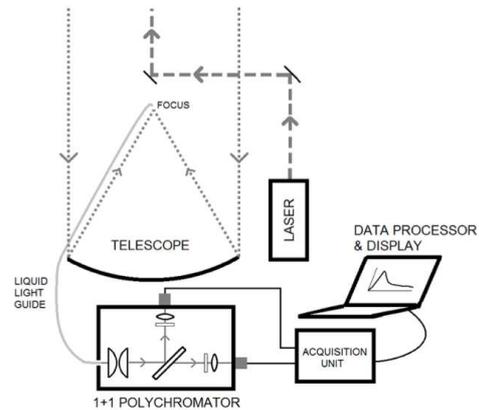}
\caption{\label{fig:lidar2}Schematic view of a possible Raman LIDAR
  for the CTA. A laser is pointed towards the atmosphere, and the
  backscattered light collected by a telescope. At the focal plane, a
  light guide transports the light to a polychromator unit which is 
  controlled and readout by an acquisition system and a data processor
  unit.}
\end{figure}

LIDAR is an acronym for {\underline LI}ght {\underline D}etection
{\underline A}nd {\underline R}anging. 
The methodology of the LIDAR technique requires the
transmission of a laser-generated light-pulse into the atmosphere (see
Fig.~\ref{fig:lidar2}). The
amount of laser-light backscattered into the field of view of an
optical receiver on the ground, is then recorded and analyzed. 
LIDARs have proven to be a powerful tool for environmental
studies. Successful characterization of the atmosphere has been made at night using these
systems~\cite{bib:inaba, bib:ansmann, bib:behrendt},
and in other fields of astronomy, the LIDAR technique has proven to be
useful for the determination of the atmospheric extinction of starlight~\cite{bib:zimmer}.
Of the various kinds of LIDARs, the so-called elastic one make only use of
the elastically backscattered light by atmospheric constituents, while
the Raman LIDARs make also use of the backscattered light from
roto-vibrational excitation of atmospheric molecules. 
Elastic LIDARs are the simplest class of LIDAR, but their backscatter
power return depends on two unknown physical quantities (the total
optical extinction and backscatter coefficients) which need to be
inferred from a single measurement. As a result various assumptions
need to be made, or boundary calibrations introduced, limiting the precision of 
the height-dependent atmospheric extinction to always worse than 20\%. The introduction of 
additional elastic channels and/or Raman (inelastic-scattering)
channels allows for
simultaneous and independent measurement of the extinction and
backscatter coefficients with no need for \emph{a priori}
assumptions~\cite{bib:ansmann}. Raman LIDARs yield a precision of the atmospheric extinction of 
better than 5\%.

The LIDAR return signal can be fully described by the LIDAR
  equation:
\begin{equation}\label{eq:lidar}
P(R,\lambda_{rec})=K\;\frac{G(R)}{R^2}\;\beta(R,\lambda_{em})\;T^\uparrow(R,\lambda_{em})\;T^\downarrow(R,\lambda_{rec})\quad,
\end{equation}
which  contains a system factor $K$ (emitted
power, pulse duration, collection area of the telescope),  a
geometrical overlap factor (overlap of the telescope field-of-view with
the laser light cone) $G(R)$,  the molecular and aerosol backscatter
coefficient $\beta(R,\lambda_{em})$ and the transmission terms
$T^{\uparrow}(R,\lambda_{em})$ and
$T^{\downarrow}(R,\lambda_{rec})$. $R$ is the atmospheric range,
i.e. the distance from the LIDAR optical receiver, and
$\lambda_{em,rec}$ are the emitted and received wavelengths.

Using the elastic  and Raman-scattered profiles, the
atmospheric aerosol extinction coefficients $\alpha^{m,p}$ ($m$ stands
for molecules and $p$ stands for particles or aerosol) can be derived
using formulas such as:
\begin{equation}\label{eq:lidar2}
\alpha^p(R,\lambda_0) =
\frac{\frac{d}{dr}\ln\left(\frac{N_{N_2}(R)}{R^2P(R,\lambda_{N_2})}\right)
  - \alpha^m(R,\lambda_0)- \alpha^m(R,\lambda_{N_2})}
{1+\left(\frac{\lambda_{0}}{\lambda_{N_2}}\right)^{\AA}},
\end{equation}
where
$\lambda_0$ is the elastic wavelength (355, 532 nm in our case) and
$\lambda_{N_2}$ is the corresponding Raman-shifted N$_2$ backscattered 
wavelengths (387, 607 nm). $N_{N_2}$ is the nitrogen number
density. Eq.~(\ref{eq:lidar2}) has only the \AA
ngstr\"om index as free parameter (if only one elastic-Raman
wavelength pair is used) and this leads to a good precision on
$\alpha^{p}$, because  over- and underestimating 
the \AA ngstr\"om index by 0.5 leads to only 5\% relative error in the
extinction factor. Hence, apart from statistical uncertainties (which
can be minimized by averaging many LIDAR return signals), results are
typically precise to about 5-10\% {\bf in each altitude bin}, and
probably even better in clear free tropospheres with only one aerosol layer.
The uncertainty generally grows with increasing optical depth of the
layer. By adding a {\bf second Raman line}, e.g. the $N_2$ line at
607~nm, the last free \AA ngstr\"om parameter becomes fixed, and
precisions of {\bf better than 5\%} can be achieved for the aerosol
extinction coefficients. The molecular extinction part needs to be
plugged in by hand using a convenient model. However, since the
molecular densities change very little, and on large time scales,  
this can be achieved by standard tools. Precisions of typically better than 2\%
are rather easy to achieve. 

%
The experience of MAGIC with an elastic LIDAR system (i.e. analyzing
only one backscatter wavelength, and no Raman lines),   has shown that
simplified reconstruction algorithms can be used to achieve
good precision of the aerosol extinction coefficients, at least within  
the range of uncertainties inherent to an elastic LIDAR \cite{bib:fruck}. 
An analog conclusion was achieved with the H.E.S.S. LIDAR: a stable
analysis algorithm was found, limited by the 30\% uncertainties of the
time and range dependent LIDAR ratio.
%
\vspace{-5mm}
\section{Raman LIDAR characterization}
\begin{figure}[h!t]
\centering
\includegraphics[width=0.90\linewidth]{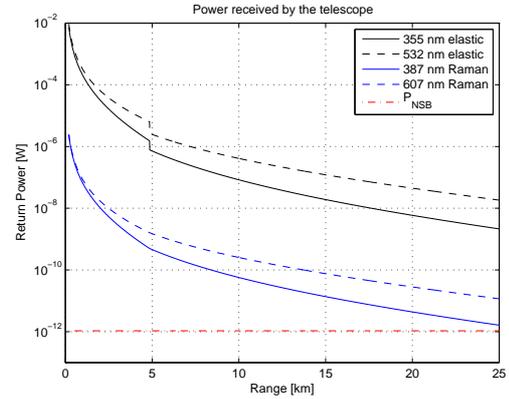} 
\caption{\label{fig:return_power}Estimated return power from the
  link-budget simulation of the Barcelona Raman LIDAR. 
The horizontal red line is the background power calculated for a typical night-sky background at an astronomic site.}
\end{figure}

Several institutes in CTA are currently designing Raman LIDAR systems:
the Institut de F\`isica d'Altes Energies (IFAE) and the Universitat
Aut\`onoma de Barcelona (UAB), located in Barcelona (Spain), the LUPM
(Laboratoire Univers et Particules de Montpellier) in Montpellier
(France) and the CEILAP (Centro de Investigaciones Laser y sus
Aplicaciones) group in Villa Martelli (Argentina)~\cite{bib:ristori}. The different
groups are designing independently the LIDAR systems with different
mechanical, optical and steering solutions. In order to assess the
performance of Raman LIDARs for CTA, we use the current baseline
design of the Barcelona LIDAR, which is also presented elsewhere in this
conference~\cite{bib:lopez}. It consists of a 1.8~m diameter parabolic mirror
equipped with a powerful Nd:YAG laser.  The outgoing laser beam at 355
and 532~nm is directed towards the telescope pointing axis in a
co-axial configuration, ensuring full near-range overlap at little
more than hundred meters. In the design of the optical readout module,  
special care has been taken to minimize signal losses throughout the
entire light collection scheme, especially for the 2 dim Raman lines
at 387 and 607~nm. For the Barcelona LIDAR, a so-called
\textit{link-budget} figure-of-merit calculation has been  
performed showing that the dimmest Raman line will be detected from a
distance of 15~km (see Fig.~\ref{fig:return_power}) with a
signal-to-noise ratio of 10 after only one minute. This short
integration time (non standard for typical LIDAR usage) is required
for CTA because the LIDAR operation should not interfere with the
experiment datataking. For example, it could be possible to perform LIDAR
campaigns entirely during the telescope repositioning time.

\vspace{-5mm}

\section{Strategies for data reconstruction}

\begin{figure}[h!t]
\centering
\includegraphics[width=0.9\linewidth]{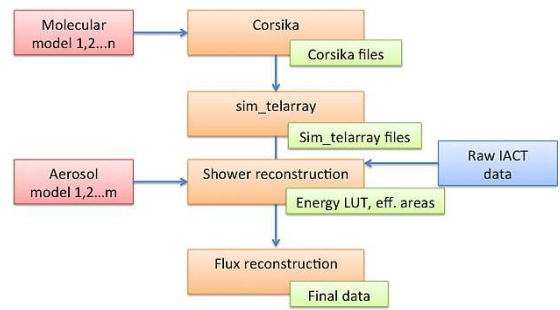}
\caption{Scheme of a possible data analysis flow in case the
  atmospheric model is used at the data level.\label{fig:correction_scheme2}} 
\end{figure}

\begin{figure}[h!t]
\centering
\includegraphics[width=0.9\linewidth]{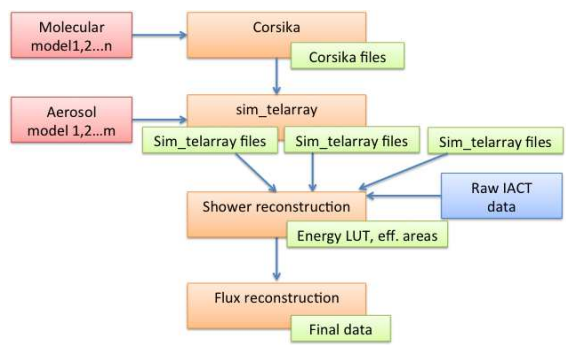}
\caption{Scheme of possible data analysis flow where the atmospheric
information is introduced in the Monte Carlo
simulations\label{fig:correction_scheme1}}  
\end{figure}

Once the atmospheric extinction profile is determined, the data taken
with the CTA observatory need to be corrected accordingly. This can be
achieved by re-calibrating the data themselves, either event-wise  
or bin-wise (Fig.~\ref{fig:correction_scheme2}), or by simulating
adapted atmospheres (Fig.~\ref{fig:correction_scheme1}). We have seen from
our simulations that data affected by enhancements of the ground layer
can be scaled rather easily up to high levels of extinction, and
probably no dedicated MC simulation is needed for each set of data.  
This is not the case anymore for (cirrus) clouds at altitudes from 6
to 12~km a.s.l., which create strong energy-dependent effects on the scaling factors. Moreover, the images are distorted depending on the location of the shower maximum, which varies even for showers of a same energy. Very high altitude layers, in turn, produce only
effects on the very low energy gamma-ray showers.  Depending on the properties of the
chosen site for CTA, still to be decided, it would probably make sense to
create $10-20$ typical atmospheric simulations within these
possibilities and interpolate between them. 

\vspace{-5mm}
\section{Complementing devices}
Apart from the Raman LIDAR, complementary devices for atmospheric
characterization and the understanding of the site climatology can be
used. A first class of devices comprises those which provide at least
some profiling of the atmosphere, such as radio sondes, profiling
microwave or infrared radiometers and differential optical absorption
spectrometers. The operating wavelengths of these devices are very
different from those of the Raman LIDAR, and precise conversion of
their results to the spectral sensitivity window of the CTA is
difficult. However, since aerosols are better visible at larger
wavelengths, profiling devices may be used to determine cloud heights
with high precision and their results may be good seeds for the Raman
LIDAR data inversion algorithm. A next class of complementary devices
contains those which measure integral parameters, such as Sun, Lunar
and stellar photometers, UV-scopes and starguiders. Integral optical
depth measurements have become world-wide standards, organized in
networks ensuring proper (cross-)calibration of all devices.  
Spectacular resolutions of better than 1\% can be obtained during the
day, about 2\% with moon, and 3\% under dark night conditions,  at
wavelength ranges starting from about 400~nm. Extrapolations to the
wavelength range between 300 and 400~nm worsens the resolution
again. The precise results from these devices can serve as important
cross-checks of the integrated differential Raman LIDAR transmission
tables. Finally, all major astronomical observatories operate cloud
detection devices, mainly all-sky cameras and/or take advantage from
national weather radars. All-sky cameras have become standardized
within the CONCAM or the TASCA networks, however important differences
in sensitivity to cirrus clouds are reported. The advantage of these
devices are their big field-of-view which allows to localize clouds
over the entire sky and makes possible online adapted scheduling of
source observations. Relatively cheap cloud sensors based on
pyrometers or thermopiles have been tested by the MAGIC collaboration
and the SITE WP of CTA. The calibration of these devices is however
complex and measurements are easily disturbed by surrounding
installations. Recent anaysis can be found in \cite{bib:daniel,bib:hahn}. 

\vspace{-5mm}
\section{Conclusions}
The CTA observatory will have two arrays of telescopes, one in
Southern and one in the Northern hemispheres
not chosen yet. 
Despite the astronomical sites are expected to have
extremely good atmospheric conditions for most of the year, with dry
clean air, the experience with H.E.S.S., MAGIC and VERITAS has shown that non-monitored atmospheric variations introduce systematic
effects on the data, which limit the energy and flux
reconstruction. With the goal of producing high-quality data for CTA,
currently
various groups are developing instruments for atmospheric monitoring
and calibration. In particular, our Monte Carlo studies have shown that
integral atmospheric transmission parameters are not sufficient to
fully characterize the atmosphere (related to the fact that
gamma-ray showers are initiated at altitudes between the strato- and troposphere) and that range-resolved measurements are
advisable. The natural solution is the use of (Raman) LIDARs, which
were described in this contribution. In addition, the use of
complementary instruments that measure integral or differential (in
altitude) atmospheric parameters is possible and envisaged. Once
retrieved the differential atmospheric transparency, different
strategies are foreseen to accurately and precisely reconstruct data,
ultimately reducing the reconstructed energy and flux uncertainties.
In addition, it would be possible to increase the duty cycle of the
telescopes by retrieving those data taken during non-optimal atmospheric
conditions which are normally discarded by standard clean-atmosphere
analysis, especially important during e.g. multi-wavelength
campaigns or target of opportunity observations. Finally, the use of atmospheric
instruments will allow for continuous weather now- and forecast.


\vspace*{3mm}
\footnotesize{{\bf Acknowledgment:}{ We gratefully acknowledge support from the agencies and organizations 
listed in this page: \url{http://www.cta-observatory.org/?q=node/22}}}

\end{document}